\documentclass[prb,preprint,aps,superscriptaddress,longbibliography]{revtex4-1}

\usepackage{graphicx, epstopdf}
\usepackage{subfigure}
\usepackage{float}
\usepackage{amsmath}
\usepackage{amsfonts}
\usepackage{amssymb}
\usepackage{bm}
\usepackage{subfigure}
\usepackage{caption}
\usepackage{hyperref}
\begin{document}

\title{Two-Dimensional Oxide Topological Insulator With Iron-Pnictide Superconductor LiFeAs Structure}

\author{Qiunan Xu}
\author{Zhida Song}
\author{Simin Nie}

\affiliation{Beijing National Laboratory for Condensed Matter Physics,
  and Institute of Physics, Chinese Academy of Sciences, Beijing
  100190, China}

\author{Hongming Weng}
\email{hmweng@iphy.ac.cn}
\affiliation{Beijing National Laboratory for Condensed Matter Physics,
  and Institute of Physics, Chinese Academy of Sciences, Beijing
  100190, China}
\affiliation{Collaborative Innovation Center of Quantum Matter,
  Beijing, China}

\author{Zhong Fang}
\affiliation{Beijing National Laboratory for Condensed Matter Physics,
  and Institute of Physics, Chinese Academy of Sciences, Beijing
  100190, China}

\affiliation{Collaborative Innovation Center of Quantum Matter,
  Beijing, China}

\author{Xi Dai}

\affiliation{Beijing National Laboratory for Condensed Matter Physics,
  and Institute of Physics, Chinese Academy of Sciences, Beijing
  100190, China}

\affiliation{Collaborative Innovation Center of Quantum Matter, Beijing, China}

\date{\today}

\begin{abstract}
By using first-principles calculations, we propose that ZrSiO can be looked as a three-dimensional (3D) oxide
weak topological insulator (TI) and its single layer is a long-sought-after 2D oxide TI with a band
gap up to 30 meV. Calculated phonon spectrum of the single layer ZrSiO indicates it is dynamically 
stable and the experimental achievements in growing oxides with atomic precision ensure that it can 
be readily synthesized. This will lead to novel devices based on TIs, the so called
``topotronic" devices, operating under room-temperature and stable when exposed in the air. Thus, 
a new field of ``topotronics" will arise. Another intriguing thing is this oxide 2D TI has the similar crystal 
structure as the well-known iron-pnictide superconductor LiFeAs. This brings great promise in realizing
the combination of superconductor and TI, paving the way to various extraordinary quantum phenomena, such as
topological superconductor and Majorana modes. We further find that there are many other isostructural compounds
hosting the similar electronic structure and forming a $WHM$-family with $W$ being Zr, Hf or La, $H$ being group
IV or group V element, and $M$ being group VI one.
\end{abstract}

\maketitle

\section{Introduction} \label{introduction}
Topological insulator (TI) has nontrivial band topology characterized by $Z_2$ topological invariant under
 time-reversal symmetry (TRS).~\cite{TIreview, TIreview-2, adv_phys} Similar to the systems with integer quantum Hall effect,
 the bulk band structure of TIs is insulating with unavoidable edge or surface states on the boundary connecting the
 valence and conduction bands. Especially for the two-dimensional (2D) TI, it can host quantum spin Hall effect (QSHE)
with 1D helical edge states, where the states moving in opposite direction have opposite spin.
Therefore, the backscattering
is prohibited as long as the scattering potential does not break TRS and such helical edge states provide a new
mechanism to realize non-dissipative electronic transportation,
 which promises potential application in low-power and multi-functional device based on TI.
From this point of view, a 2D TI with large band gap, chemically stable under ambient condition exposed in the air, easy to prepare and consisting of
cheap and nontoxic elements is highly desired.~\cite{weng_MXene, MRS_weng:9383312} The study of TI based quantum devices would foster a field of
``topotronics". Such 2D TI is also important and crucial to the realization of topological superconductivity and Majorana modes through
proximity effect.~\cite{TS_PRL2008,TIreview-2}

However, up to now there are only a few systems, which have been proved to be 2D TI,~\cite{ando_topological_2013, MRS_weng:9383312} such as the quantum well of HgTe/CdTe~\cite{bernevig_science_quantum_2006, konig_quantum_2007} and InAs/GaSb.~\cite{PhysRevLett.107.136603} Both of above systems require very sophisticated
and expensive MBE growth and only show QSHE at ultra-low temperature. These extreme requirements have obstructed further studies and possible applications
of 2D TIs. The theoretical efforts in finding and designing 2D TI candidates keep going on and some of them are listed here: silicene~\cite{liu_quantum_2011}, chemically decorated single
layer honeycomb lattice of Sn,~\cite{PhysRevLett.111.136804} Ge~\cite{PhysRevB.89.115429} and Bi or Sb,~\cite{BiX_Yao2014, Nano_QSHE, 2D_IATI}, buckled BiF~\cite{XiangSQ},
single layer ZrTe$_5$, HfTe$_5$~\cite{weng_transition-metal_2014} and Bi$_4$Br$_4$,~\cite{zhou_large-gap_2014} transition-metal dichalcogenide (TMD)
in 1T'~\cite{qian2014quantum} and haeckelite~\cite{ smnie2015, song2015, yanbh, daiying} structure. But none of them has been confirmed experimentally. Recently,
Weng {\it et al} proposed a large band gap 2D TI in oxygen functionalized MXene,~\cite{weng_MXene}
which brings the hope in realizing TI in oxygen contained compounds, which are naturally antioxidant and stable upon exposed in the air. This also stimulated
the effort in this paper in searching for new oxide TIs.

In this paper, by using first-principles calculations we demonstrate that the oxide material ZrSiO is a weak TI and a single layer of it
is a 2D TI with achievable band gap up to 30 meV.~\cite{3dTI1, 3dTI2, 3dTI3}
 We have calculated its edge states and analyzed the physical mechanism for band inversion. We have found many other
isostructural compounds with formula of $WHM$ ($W$=Zr, Hf or La, $H$=Si, Ge, Sn or Sb and $M$=O, S, Se and Te) possessing the similar electronic structure.

\section{Computational Details} \label{method}
We have employed the Vienna $ab~ initio$ simulation package (VASP)\cite{kresse1996_1,kresse1996_2} for most of the density functional theory (DFT) based first-principles calculations.
Exchange-correlation potential is treated within the generalized gradient approximation (GGA) of Perdew-Burke-Ernzerhof type.\cite{Perdew1996}
Spin-orbit coupling (SOC) is taken into account self-consistently. The cut-off energy for plane wave expansion is 500 eV and the k-point sampling grid in the self-consistent process was 12$\times$12$\times$3 for 3D case and 12$\times$12$\times$1 for the single layered structure.
The crystal structures have been fully relaxed until the residual forces on each atom is less than 0.001 eV/\AA. A vacuum around 16~\AA ~between
layers was used to minimize the interactions between the layer and its periodic images.  The possible underestimation of band gap within GGA is checked
by non-local Heyd-Scuseria-Ernzerhof (HSE06) hybrid functional\cite{heyd2003hybrid, heyd2006hybrid} calculation. To explore the edge states, we
build a tight-binding model for a slab around 120 unit-cell thick with edges along $a$ or $b$ lattice.~\cite{MRS_weng:9383312, adv_phys} The maximally localized
Wannier functions (MLWF)~\cite{marzari1997,souza2001} of $d$ orbitals of $W$ and $p$ orbitals of $H$ are generated as the basis set by
using the software package OpenMX.~\cite{openmx, weng_mlwf}

\begin{figure}
\includegraphics[width=0.8\textwidth]{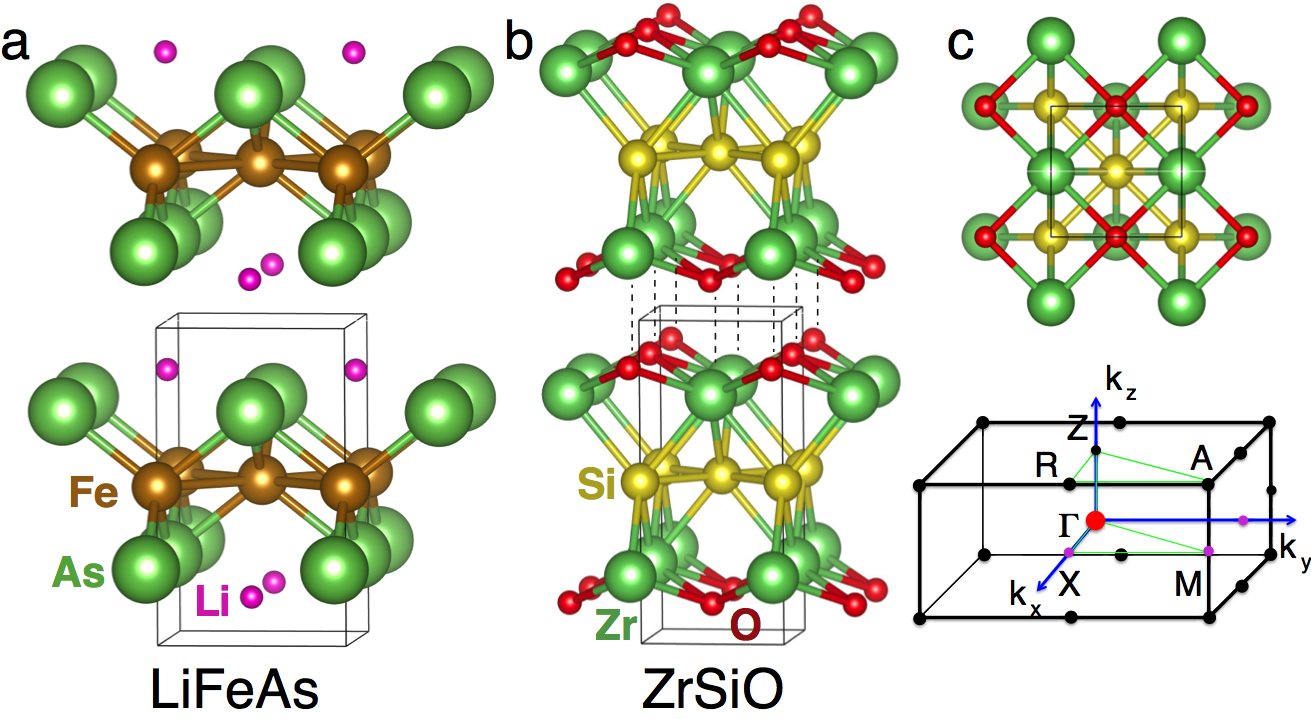}
\caption{(Color online) Crystal structure of (a) LiFeAs (b) ZrSiO. (c) Top view of single layer ZrSiO and 3D Brillouin zone with high symmetrical crystal momenta indicated. The weak bonds between Si-O have been indicated by dashed lines, which form a natural cleavage plane to get 2D single layer ZrSiO.}
\label{crystructure}
\end{figure}

\section{Results and Discussion} \label{Results}
{\it Crystal Structure.}  The bulk ZrSiO belongs to a big family of ZrSiS- (or PbFCl-) type compounds~\cite{ZrSiO_crystr} with chemical formula of
$WHM$ ($W$=transition metal or rare earth element; $H$, $M$=main group elements).
There are over 200 members in this family,~\cite{ZrSiS_type} and the crystal structure in Fig.~\ref{crystructure} can be looked as stacking of five square nets of $W$, $H$ and $M$ in sequence of [... $M$-$W$-$H$-$W$-$M$ ...] along $c$ axis.
The relatively weak $M$-$W$ bonding connecting two neighboring slabs show in Fig.~\ref{crystructure} leads to quasi layered structure of them.
In fact, it was thought that the cross over from 3D to 2D in ZrSi$M$ ($M$=O, S, Se and Te) happens between $M$=Se and Te.~\cite{ZrSiTe_2d} Nevertheless, this gives out a natural cleavage plane for getting 2D layer of these compounds and we will show the band dispersion along $c$ is quite narrow than that of in-plane for several members in this family.
It is noticed that the isostructural pnictide NbSiAs has been found to be superconductor.~\cite{NbSiAs_supercond} The intensively studied iron-pnictide superconductor LiFeAs also has the iso-structure but with the cation and anion sites
being switched to form the so called anti-PbFCl (or anti-ZrSiS) type structure.~\cite{LiFeAs} The space group of ZrSiO is P4/nmm (No. 129). The theoretically optimized lattice constants are $a$=$b$=3.316~\AA ~and $c$=7.354~\AA, being underestimated by about 6-7\% compared with experimental values $a$=$b$=3.52~\AA ~and $c$=7.93~\AA.~\cite{ZrSiO_crystr} The is within the common error in lattice
constant estimation from DFT calculation within GGA. The optimized Wyckoff site of atoms are Zr (0.25, 0.25, 0.308), Si (0.75, 0.25, 0) and O (0.25, 0.25, 0.6112), also consistent with the
experimental values. The general features of the band structure from the optimized crystal structure and the experimental one are quite similar. To be consistent
in the description of the main text, the theoretically optimized crystal structure is used in the following discussion.

\begin{figure}
\includegraphics[width=0.8\textwidth]{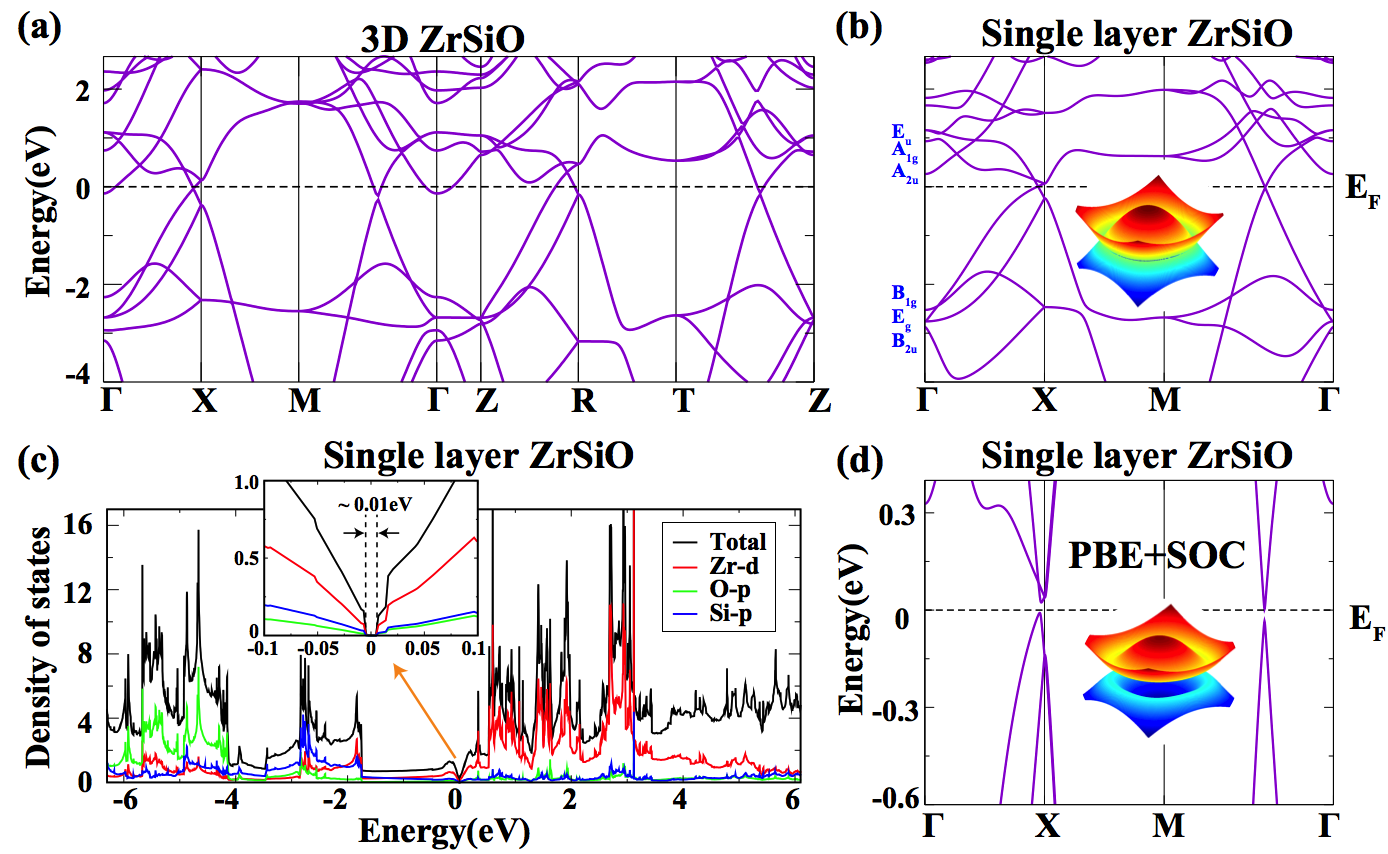}
\caption{(Color online) Band structure for (a) bulk ZrSiO within GGA and single layer ZrSiO within (b) GGA and (d) GGA+SOC. The inset in (b) indicates the Node-Line due to band inversion around Fermi level protected by in-plane mirror symmetry. The inset in (c) indicates the gap-opening
in Node-Line when SOC is further included.}
\label{bands2}
\end{figure}

{\it Band structure of 3D ZrSiO.} The band structure of 3D ZrSiO is shown in Fig.~\ref{bands2}(a). There are two distinct features. One is the apparent 
band crossing points around the Fermi level indicating the existence of band inversion as found in many other materials~\cite{weng_transition-metal_2014, smnie2015} with nontrivial topological band structures.~\cite{MRS_weng:9383312, adv_phys}
The second feature is its band dispersion in $k_z$=0 plane ($\Gamma$-X-M-$\Gamma$) resembles that in $k_z$=$\pi$ plane (Z-R-A-Z) and there is no band 
inversion happen along $\Gamma$-Z direction. This means that its electronic structure is essentially more 2D like than 3D. The further group theory analysis 
indicates that the bands crossing in both $k_z$=$\pi$ and $0$ planes are protected
by glide mirror perpendicular to the $z$-axis because the two bands forming the crossing points belong to different eigenvalues of that glide mirror.
As discussed in Ref.~\onlinecite{TaAs_Weng, allcarbon_nodeLine2014, Cu3NPd, Cu3NPdKane},
these band crossing points form a closed node-line circle in the $k_z$=0 and $k_z$=$\pi$ plane, respectively, when SOC is not considered. Including SOC into calculation
opens a gap along the node-lines turning 3D ZrSiO into a system with finite gap at all $k$ points but compensated electron and hole pockets at the Fermi level. As we have discussed
in Ref.~\onlinecite{allcarbon_nodeLine2014, Cu3NPd}, the crossing points on the node-line are not exactly on the same energy level
due to the  particle-hole asymmetry.  There will be a finite global band gap instead of compensated electron-hole pockets.
Since the material has inversion center, we can determine its $Z_2$ invariances through the production of the parities at the time reversal invariant momenta (TRIM),
which gives (0; 001) indicating that it is a weak TI  if the band gap opening  on the node-lines due to SOC is larger enough to open the indirect gap.
~\cite{3dTI1, 3dTI2} The above results also indicate that 3D ZrSiO can be viewed as the
stacking of 2D TI along $c$-axis,~\cite{3dTI1,3dTI2} similar to ZrTe$_5$~\cite{weng_transition-metal_2014} and Bi$_4$Br$_4$.~\cite{zhou_large-gap_2014}

\begin{figure}
\includegraphics[width=0.8\textwidth]{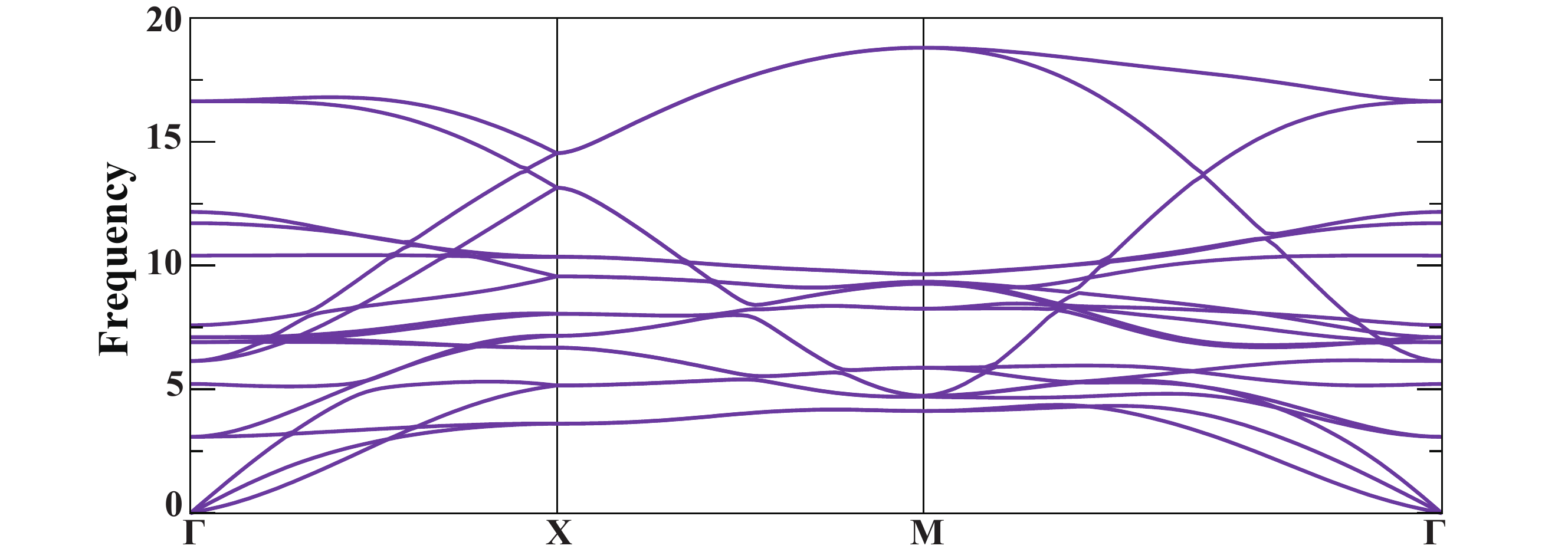}
\caption{(Color online) The phonon spectrum of optimized single layer ZrSiO. 
}
\label{phonon}
\end{figure}

\begin{figure}
\includegraphics[width=0.8\textwidth]{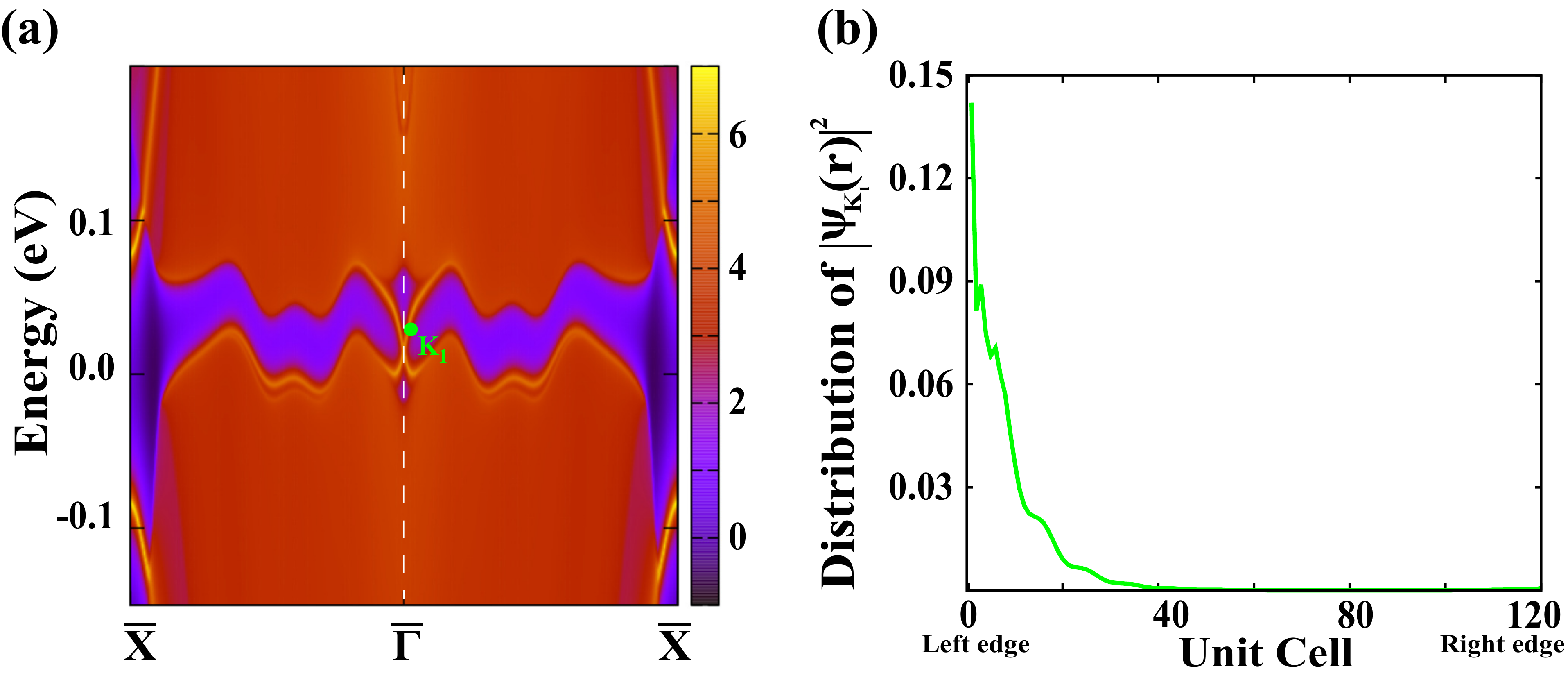}
\caption{(Color online) The edge state of single layer ZrSiO along $a$ or $b$ lattice vector. (a) The Dirac cone like edge state connects the bulk valence and conduction bands. (b) The wave function at the sampled $k$ in (a) on the edge state shows its decay into bulk as function of number of unit cell.
}
\label{edge_state}
\end{figure}

{\it 2D TI of Single layer ZrSiO.} As discussed above, we take a single layer of ZrSiO by breaking the weak Si-O bonds. In order to get the binding energy 
between layers of ZrSiO, we calculate the total energies of 3D and single layer ZrSiO with the same setting of parameters in calculations. The binding energy 
is about 1.166 or 1.503 eV as shown in Table~\ref{table:ZrHM} and ~\ref{table:ZrSiM} calculated within GGA and LDA, respectively. It is much bigger 
than the interlayer binding energy of graphite and ZrTe$_5$~\cite{weng_transition-metal_2014} and Bi$_4$Br$_4$.~\cite{zhou_large-gap_2014}. In this point 
of view, the interlayer interaction in ZrSiO is weak covalent bonding rather than van der Waals interaction. 

On the other hand, the band structure of single layer ZrSiO essentially has the same band topology as that in any plane perpendicular to $k_z$ axis in 3D ZrSiO. 
The only difference is in the strength of particle-hole asymmetry. On considering these and the experimental technique of well controlled layer-by-layer growth 
of oxides~\cite{LAOSTO}, we think the single layer ZrSiO is readily synthesized. We perform full geometrical optimization of the free standing single layer ZrSiO. The cell 
parameters $a$ and $b$ have been relaxed to $a$=$b$=3.225~\AA. In the direction $c$, the atoms Zr move close to Si layer by about 0.0251~\AA, and atoms 
O move close to Si layer by about 0.004~\AA. Such small relaxation can also be ascribed to its quasi-2D feature. The phonon spectrum of the fully relaxed structure
is shown in Fig.~\ref{phonon}. It has no imaginary frequency and indicates the dynamical stability of single layer ZrSiO.

The band structures shown in Fig.~\ref{bands2} (b)-(d) are for fully optimized 2D ZrSiO. As shown in Fig.~\ref{bands2} (b), the two Dirac points on 
the paths $\Gamma$-X and $\Gamma$-M are nearly at the same energy level close to Fermi energy. 
When SOC is further considered, a global band gap around 0.01 eV is obtained as indicated by the total density of states (DOS) in Fig.~\ref{bands2} (c) 
and the band structure in Fig.~\ref{bands2} (d). The insets in Fig.~\ref{bands2} (b) and (d) clear show the gap opening by SOC along the node-line circle.
We find that by slightly compressing the thickness of the slab or equivalently, expanding the in-plane lattice constants would decrease the particle-hole asymmetry
and change the size of band gap. For example, taking the experimental in-plane lattice constant and compressing the ZrSiO layer by 4\% would lead to a 
global band gap about 30 meV. 

Since it is central symmetric, the $Z_2$ number 1 is determined by counting the parity of all occupied states at four time reversal invariant momenta.~\cite{3dTI1,3dTI2}
This confirms that the single slab of ZrSiO is a 2D TI. Its edge state along $a$ or $b$ lattice vector is shown in Fig.~\ref{edge_state}, which is 
calculated using the Green's function method~\cite{MRS_weng:9383312, adv_phys} based on the tight-binding hamiltonian in basis of Wannier 
functions of Zr $d$ and Si $p$ orbitals.~\cite{openmx, weng_mlwf} The edge states are characteristic Dirac cone like. We have also studied the 
decaying of edge state into the interior of 2D ZrSiO. As shown in Fig.~\ref{edge_state} (b), it decays quite slow compared with the
surface states of 3D TIs of Bi$_2$Se$_3$ or Bi$_2$Te$_3$. Therefore, a quite thick slab is needed, more than 80-unit-cell thick along 
$a$ or $b$ lattice, to get gapless Dirac cone along its edge.

\begin{figure}
\includegraphics[width=0.8\textwidth]{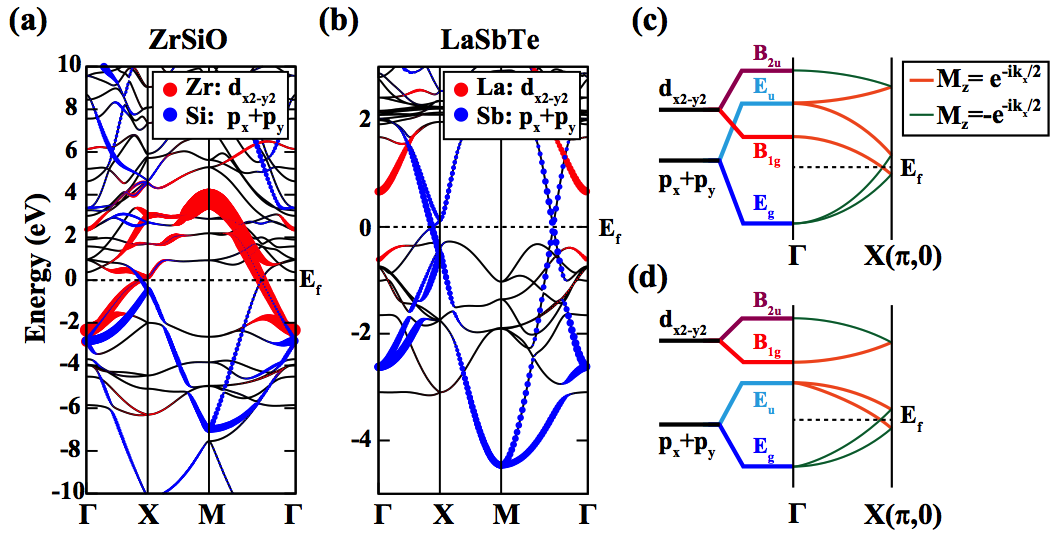}
\caption{(Color online) The band inversion mechanism in single layer ZrSiO. (a) The fatted band structure for Zr $d_{x^2-y^2}$ and Si $p_x+p_y$ orbitals. 
(b) The fatted band structure for La 5$d_{x^2-y^2}$ and Sb $p_x+p_y$ orbitals. (c) and (d) are the band inversion mechanism with and without hybridization 
between $d_{x^2-y^2}$ and $p_x+p_y$ orbitals, corresponding to ZrSiO and LaSbTe, respectively.
}
\label{bandinv}
\end{figure}

\begin{figure}
\includegraphics[width=0.8\textwidth]{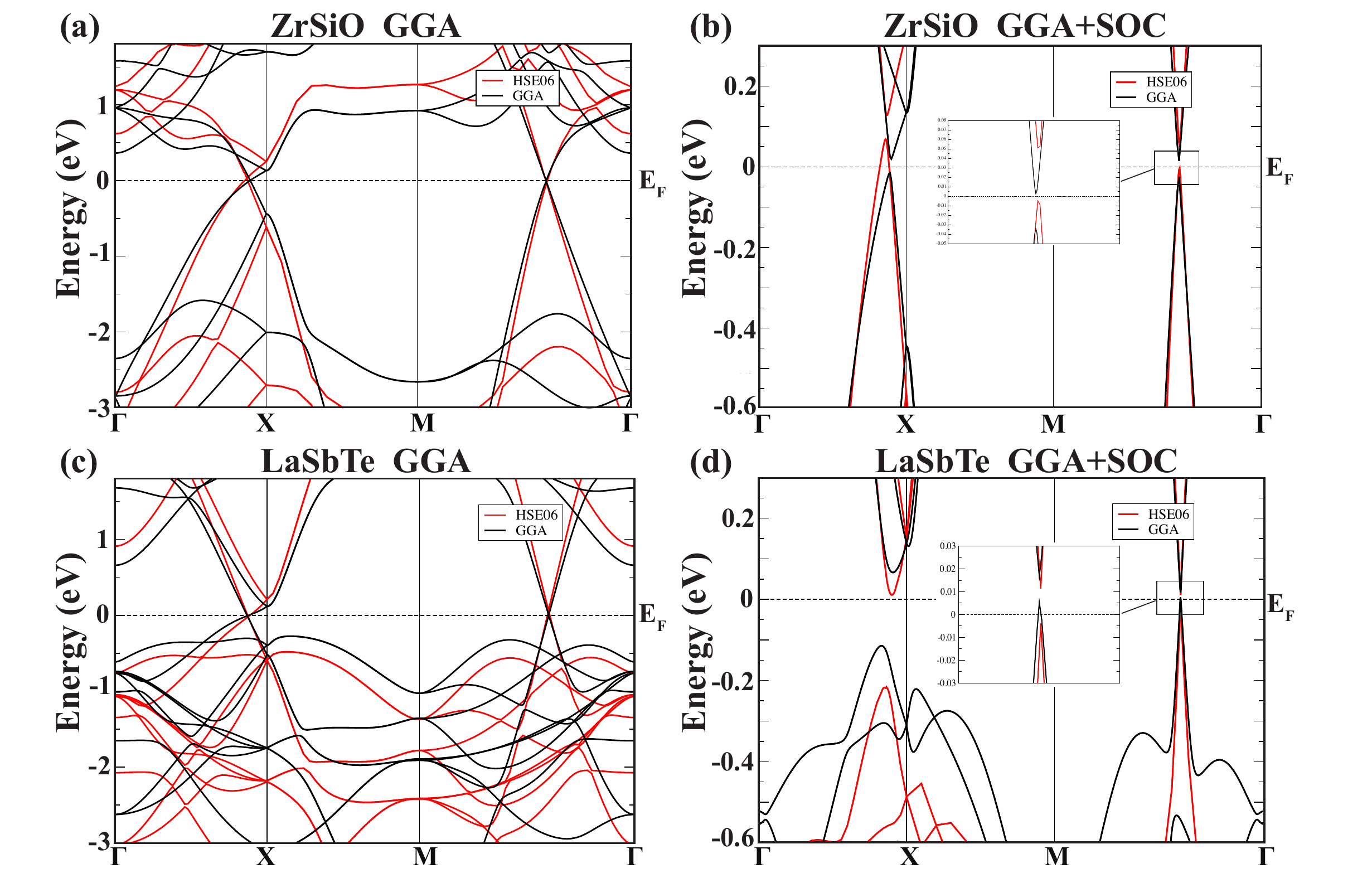}
\caption{(Color online) The band structures of (a) ZrSiO without SOC, (b) ZrSiO with SOC, (c) LaSbTe without SOC and (d) LaSbTe with SOC, calculated within GGA (black lines) and HSE06 (red lines). 
}
\label{bandhse}
\end{figure}

{\it Band Inversion Mechanism.} The crystal field splits five Zr $d$ orbitals into $d_{x^2-y^2}$, $d_{xy}$, $d_{z^2}$, and $d_{yz}+d_{xz}$ manifolds and three Si $p$ orbitals
into $p_z$ and $p_x+p_y$. The DOS plot and the fatted bands in Fig.~\ref{bandinv}(a) indicate that the bands involved into band inversion around Fermi level are mainly composed
of Zr $d_{x^2-y^2}$ and Si $p_x+p_y$ orbitals. In each unit cell, there are two nonequivalent Zr and Si atoms. The band inversion mechanism can be understood as following. Firstly,
the degenerate atomic orbitals form the bonding and anti-bonding states, with the splitting determined by the effective hopping parameters of $t_d$ and $t_p$. The bonding and anti-bonding
states from $d_{x^2-y^2}$ orbitals constitute the B$_{2u}$ (odd parity) and B$_{1g}$ (even parity) representation of D$_{4h}$, the little group at $\Gamma$. Those from degenerate
$p_x+p_y$ orbitals construct the 2D representation E$_u$ and E$_g$ of D$_{4h}$. It is noted that the $d_{xz}+d_{yz}$ of Zr atoms form the same 2D representation as $p_x+p_y$ orbitals
and they are heavily hybridized. Secondly, the banding effect along $\Gamma$-X leading to band broadening and band inversion. As we note that both the glide mirror
\{$M_z | $($\frac{1}{2}$, $\frac{1}{2}$,0)\} (with $ab$ plane as mirror plane) and the inversion symmetry $P$ anti-commute with the screw rotation \{$C_{2x}|$($\frac{1}{2}$,0,0)\} at
three TRIM points at the zone boundary X, Y and M.
As a consequence, similar to the situation in ZrTe$_5$ we can prove that all the states at those TRIM points at the zone boundary
are doubly degenerate with opposite glide mirror eigenvalue and parity, which leads to a unique property that the $Z_2$ invariance
here is determined by the order of states at $\Gamma$ point point, because all the contributions from the parity of the states at
the zone boundary TRIM points always cancels each other.
Since the states at X must be paired with glide mirror eigenstates with opposite eigenvalues $\pm e^{ik_x/2}$, as shown in Fig.~\ref{bandinv} (c) and (d), when the bands evolves
from $\Gamma$ to X, there are unavoidable band crossings for two typical order of states at $\Gamma$ from two different such materials (ZrSiO and LaSbTe).
As discussed in the previous section, these band crossings are protected by the glide mirror symmetry
and form the node-line inside of the mirror plane as shown in Fig.~\ref{bands2} (b). With SOC included, the band gap will open in ZrSiO and its $Z_2$ invariance can be
determined by the states at $\Gamma$ point to be odd.
The fitted band structure
from the tight-binding model with basis of $d_{x^2-y^2}$ and $p_x+p_y$ orbitals can well reproduce the one from first-principles calculation as shown in Fig.~\ref{bandinv} (c). The relative
energy level of $d_{x^2-y^2}$, $p_x+p_y$ and Fermi level E$_f$ determines the band topology. With the same number of electrons as ZrSiO, it is trivial insulator if $d_{x^2-y^2}$ bands
are fully occupied as shown in Fig.~\ref{bandinv} (c). However, if $d_{x^2-y^2}$ are higher than $p_x+p_y$ orbitals, Fermi level will pass through the band crossing point and it becomes
a 2D TI with SOC included. We find that the isostructural LaSbTe~\cite{ZrSiS_type} with the same number of electrons exactly follows this picture and its band structure is shown in Fig.~\ref{bandinv} (b).
The La 5$d$ orbitals are full empty and have one less valence electron than Zr or Hf, while the Sb has one more valence electron than Si and other group IV element.
The single layer LaSbTe is a 2D TI with global band gap around 0.011 eV with SOC included. The band inversion and the band gap of ZrSiO and LaSbTe are further checked 
by HSE06 calculations in Fig.~\ref{bandhse}. We find that in both cases the band inversion is well kept. The particle-hole asymmetry in ZrSiO is modified a little
and the global band gap is missing. For LaSbTe, the band gap is slightly increased to 0.015 eV.

\begin{figure}
\includegraphics[width=0.8\textwidth]{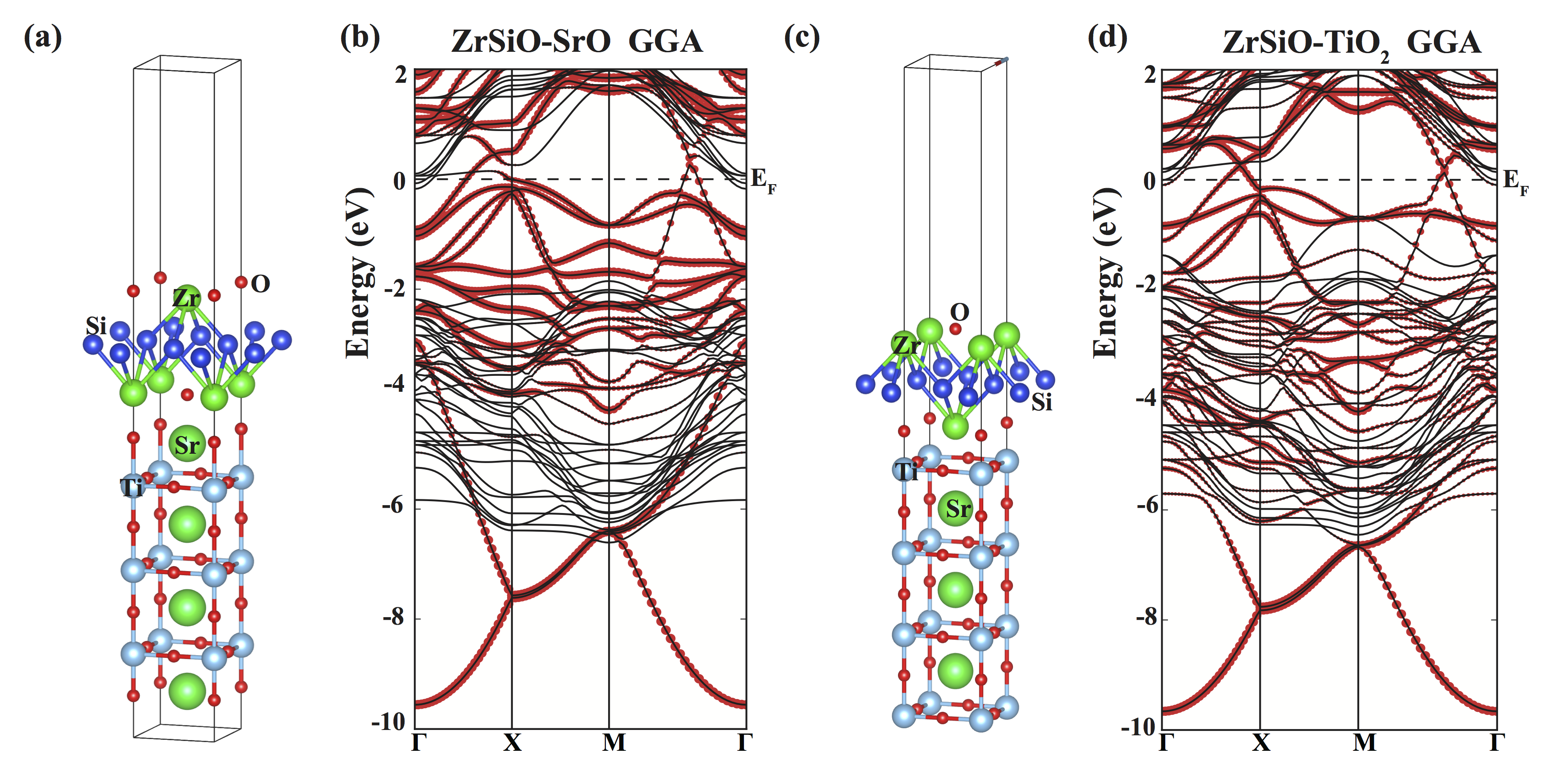}
\caption{(Color online) The geometrical structure and band structure of single layer ZrSiO on (a)-(b) SrO terminated surface and (c)-(d) TiO$_2$ terminated surface of SrTiO$_3$. The red circled lines 
are the fatted bands with the circle size proportional to the weight on single layer of ZrSiO.
}
\label{substrate}
\end{figure}

{\it Single layer ZrSiO on substrate SrTiO$_3$.} In fact, the single layer material should be supported by substrate in experiment. To check the influence of substrate on the electronic structure of ZrSiO, 
we put a single layer of ZrSiO on commonly used substrate SrTiO$_3$.  There are two types of top surface for SrTiO$_3$, namely, the SrO terminated surface and the TiO2$_2$ terminated one, as shown 
in Fig.~\ref{substrate} (a) and (c), respectively. Three unit cells of SrTiO$_3$ are taken as substrate. When ZrTiO is put on each surface, the cations are put on top of anions to minimize the total energy
and the $a$, $b$ lattice constants are expanded to fit the SrTiO$_3$ substrate. The initial distance of them is about half of cell parameter of SrTiO$_3$. There is a vacuum layer about 13~\AA ~is used. 
The three-layer atoms of SrTiO$_3$ at the interface and all atoms of single layer ZrSiO are fully relaxed. The geometrical structure has nearly no change. The fatted bands with weight on single layer 
of ZrSiO are shown Fig.~\ref{substrate} (b) and (d). We find that the most features of free standing ZrSiO is well kept, especially the band crossings around Fermi level are well reproduced inside of 
the band gap of SrTiO$_3$ substrate. On considering the underestimation of SrTiO$_3$ band gap by about 2.0 eV within GGA, we believe that SrTiO$_3$ is quite suitable for supporting ZrSiO.

\begin{figure}
\includegraphics[width=0.9\textwidth]{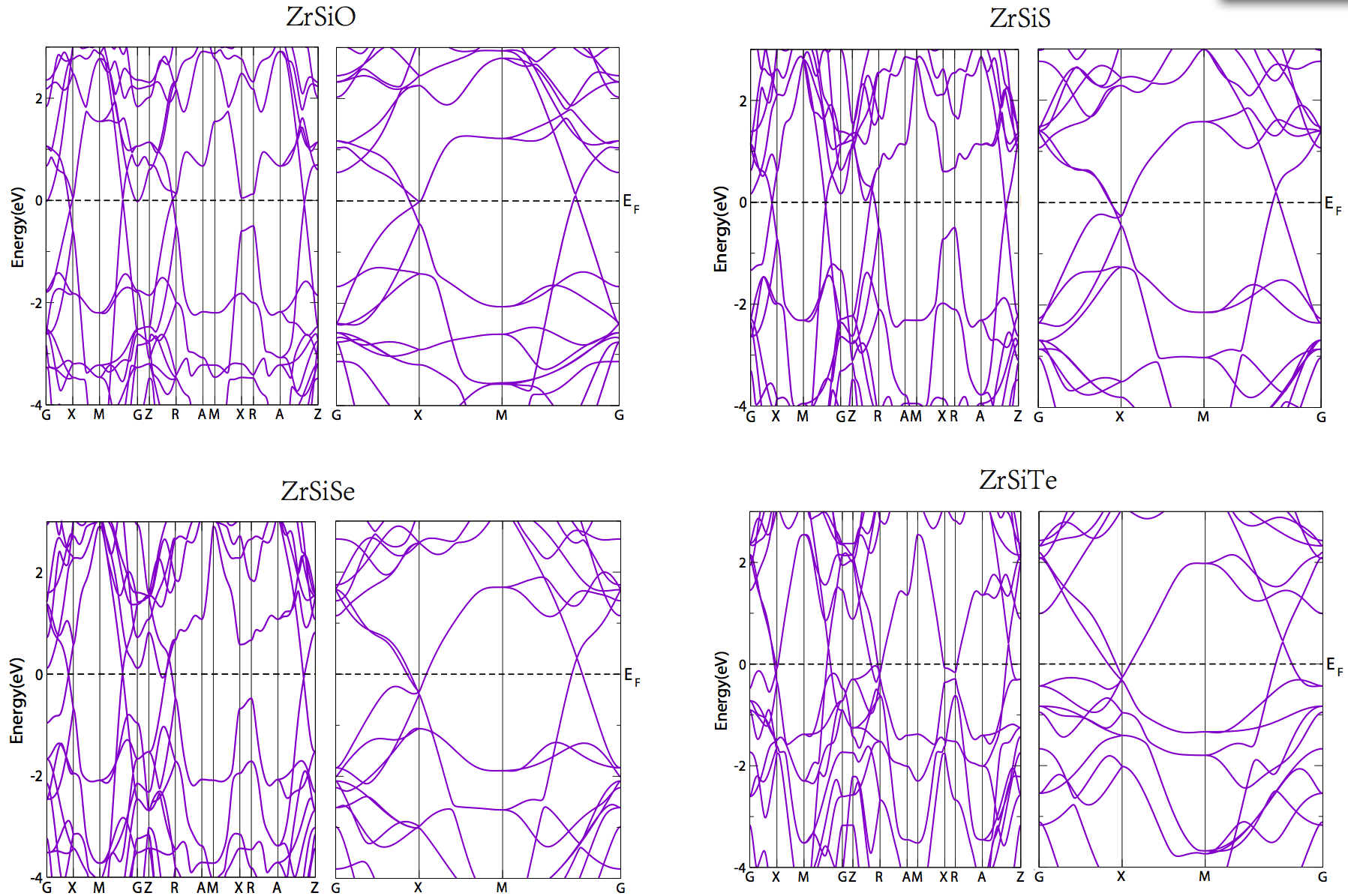}
\caption{(Color online) The band structure for (left) 3D and (right) 2D ZrSi$M$ with $M$=O, S, Se and Te. The experimental crystal structure of them are used.
}
\label{zrsio}
\end{figure}

\begin{figure}
\includegraphics[width=0.9\textwidth]{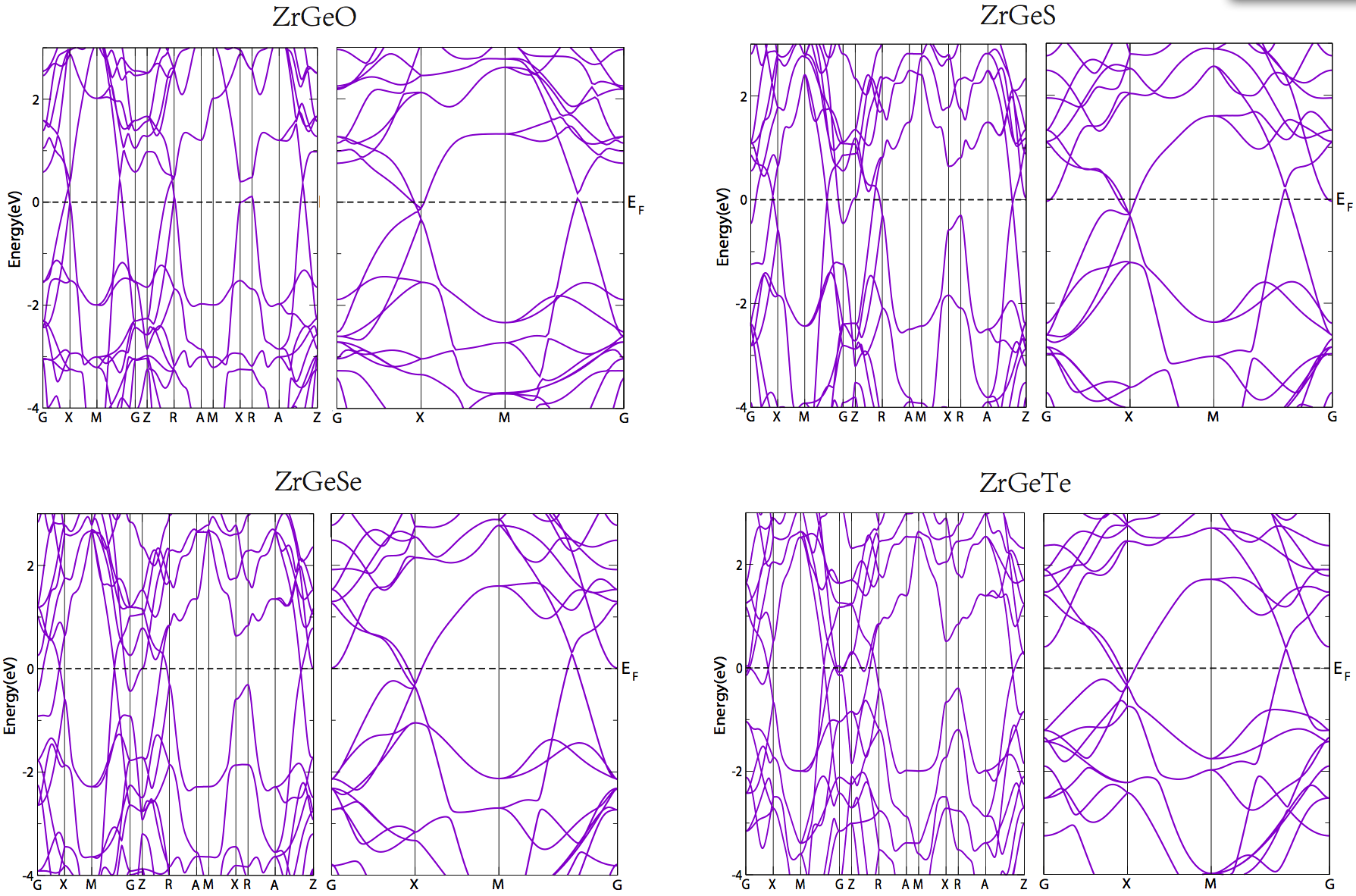}
\caption{(Color online) The band structure for (left) 3D and (right) 2D ZrGe$M$ with $M$=O, S, Se and Te. Except $M$=O case, the others are calculated with experimental structure.
The experimental crystal structure of ZrSiO is used for ZrGeO.
}
\label{zrgeo}
\end{figure}

\begin{figure}
\includegraphics[width=0.9\textwidth]{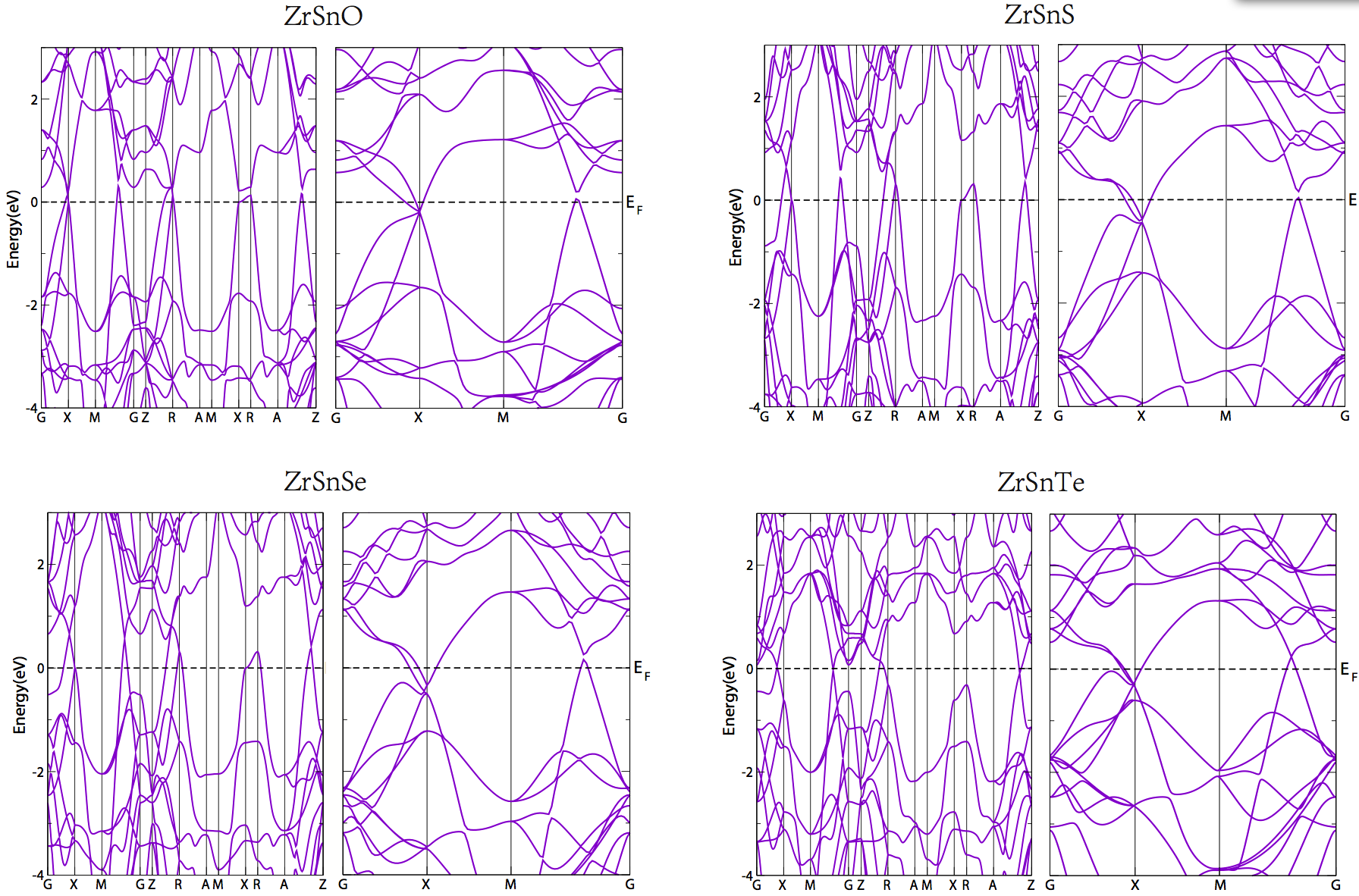}
\caption{(Color online) The band structure for (left) 3D and (right) 2D ZrSn$M$ with $M$=O, S, Se and Te. The experimental crystal structure of ZrSiO, ZrGeS and ZrGeSe are
used from ZrSnO, ZrSnS and ZrSnSe, respectively.
}
\label{zrsno}
\end{figure}

\begin{figure}
\includegraphics[width=0.9\textwidth]{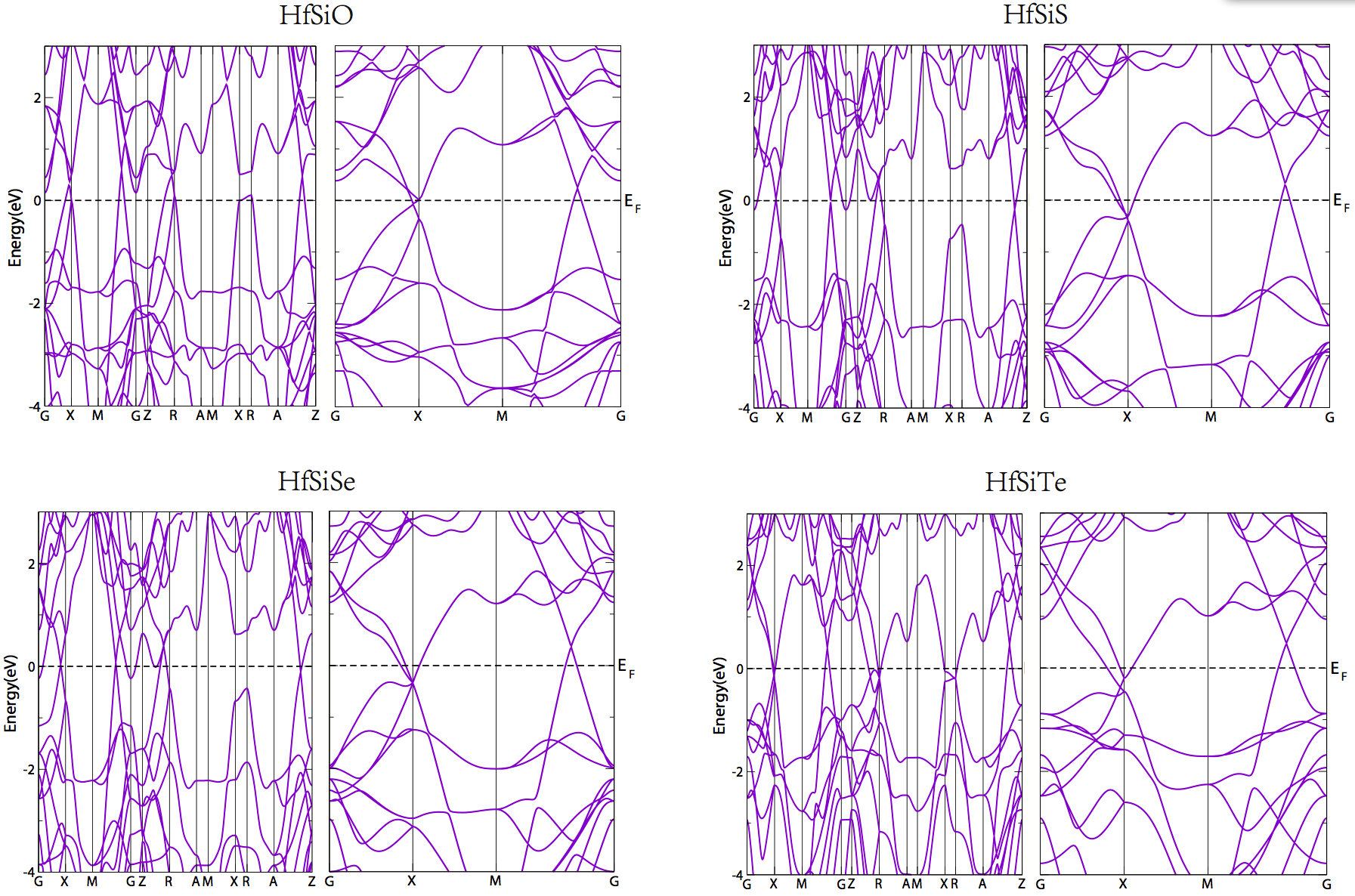}
\caption{(Color online) The band structure for (left) 3D and (right) 2D HfSi$M$ with $M$=O, S, Se and Te. The experimental crystal structure of HfSiS is used for HfSiO. For others
the experimental crystal structure is used.
}
\label{hfsio}
\end{figure}

\begin{figure}
\includegraphics[width=0.9\textwidth]{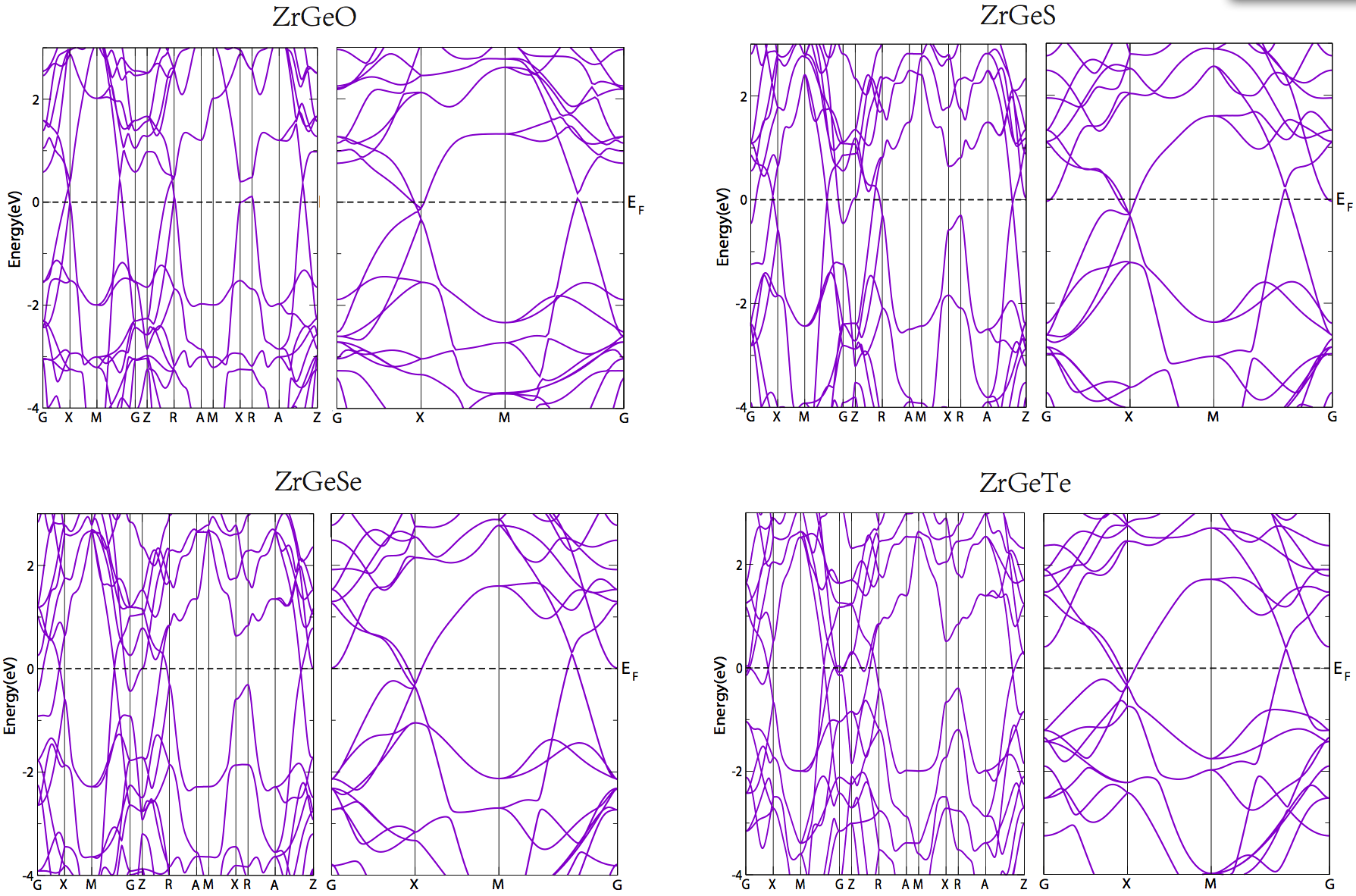}
\caption{(Color online) The band structure for (left) 3D and (right) 2D HfGe$M$ with $M$=O, S, Se and Te. The experimental crystal structure of HfSiS is used for HfGeO. For others
the experimental crystal structure is used.
}
\label{hfgeo}
\end{figure}

\begin{figure}
\includegraphics[width=0.9\textwidth]{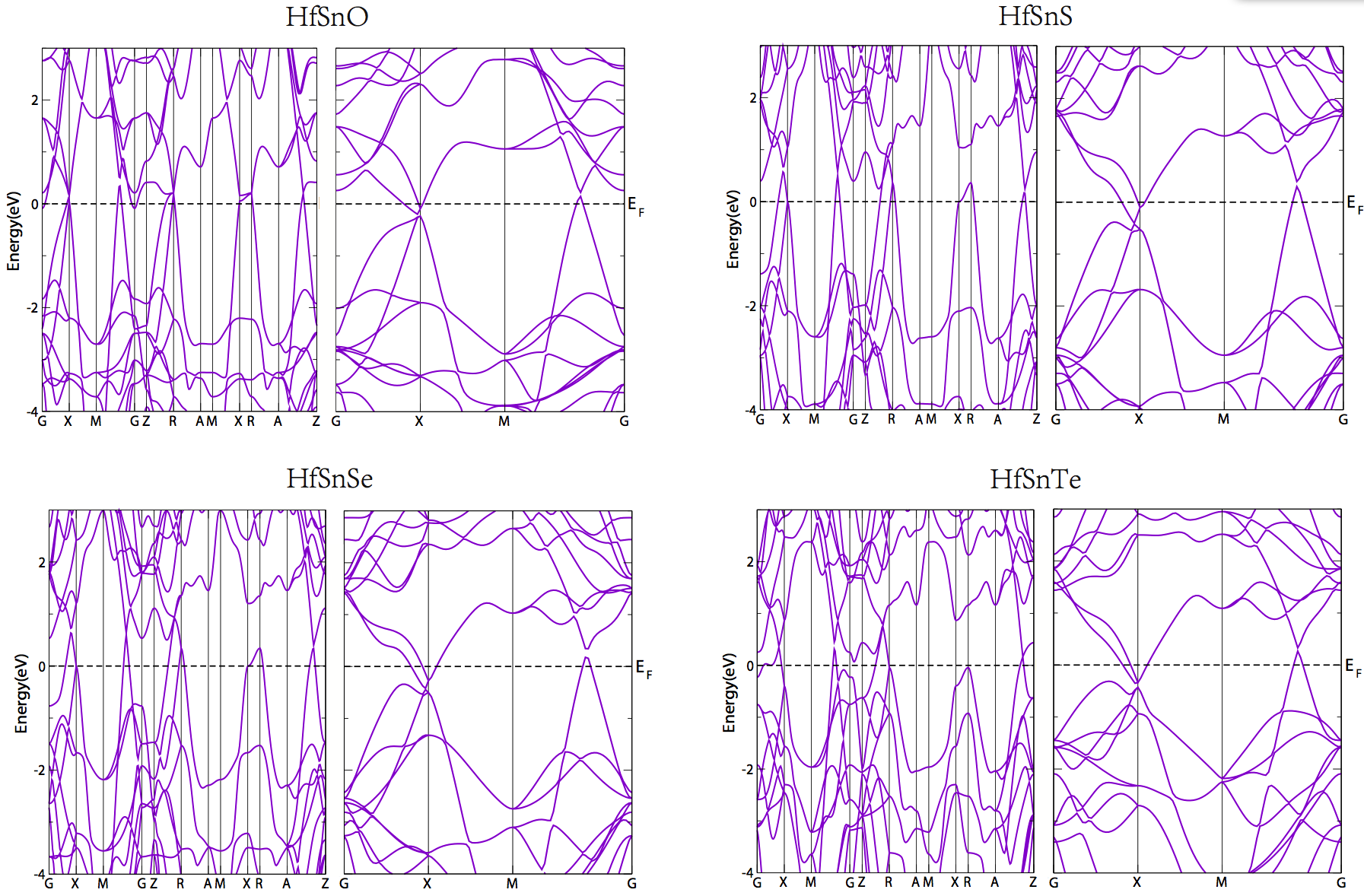}
\caption{(Color online) The band structure for (left) 3D and (right) 2D HfSn$M$ with $M$=O, S, Se and Te. The experimental crystal structure of HfSiS is used for HfSnO and HfSnS. That
of HfGeSe and HfGeTe is used for HfSnSe and HfSnTe, respectively.
}
\label{hfsno}
\end{figure}

{\it The $WHM$ family.} We have further calculated the other isostructural compounds $WHM$ with $W$=Zr, Hf, $H$=Si, Ge, Sn and $M$=O, S, Se and Te. The band structures for both the 
3D and single layer 2D $WHM$ have been shown in Figs.~\ref{zrsio},~\ref{zrgeo},~\ref{zrsno},~\ref{hfsio},~\ref{hfgeo}, and ~\ref{hfsno}.
All of these bands show the similar band structure around the Fermi level. The difference is in the strength of particle-hole asymmetry, which brings most of them are semimetals with 
overlapped electron-hole pockets.
As we have discussed above, the expansion in $a$, $b$ lattice constant and compression of the slab thickness can be helpful in decreasing the asymmetry and thus leads to the global band gap.
From these figures, we find the particle-hole asymmetry increases as $M$ changes from O to Te. At the same time, the binding energy between layers with GGA in table ~\ref{table:ZrHM} 
and ~\ref{table:HfHM} decreases when $M$ changes from O to Te. It is also validated by the example of ZrSi$M$ with LDA in Table.~\ref{table:ZrSiM}. Therefore, the further detailed 
works on the $WHM$ family are necessary and left for successive paper.

\section{Conclusions}
Based on the first-principles calculations, we have predicted that 3D ZrSiO is a weak TI and its single layer ZrSiO is an oxide 2D TI with band gap around 30 meV. The other members in this 
family noted as $WHM$ with $W$=Zr, Hf, $H$=Si, Ge, Sn and $M$=O, S, Se and Te have the similar band structure and are most probably 2D TIs. The isoelectronic compound LaSbTe has 
been shown share the similar band structure and share the same underlying physics in band inversion.

\section{Acknowledgments}
We acknowledge the supports from National Natural Science Foundation of China (Grant Nos. 11274359 and 11422428),
the National 973 program of China (Grant Nos. 2011CBA00108 and 2013CB921700) and the ``Strategic Priority Research Program (B)"
of the Chinese Academy of Sciences (Grant No. XDB07020100). Partial of the calculations were preformed on TianHe-1(A), the National Supercomputer Center in Tianjin, China.

\bibliography{ZrSiO_ref}

\newpage

\begin{table}[h]
\caption{Binding energy of Zr$HM$ ($H$=Si, Ge, Sn and $M$=O, S, Se and Te) with GGA}
\centering
\begin{tabular}{c ccc}
\hline\hline
              &   Bulk Energy (eV) &   Single layer Energy (eV) &   Binding Energy (eV) \\
\hline
ZrSiO          &   -46.130021 &   -44.963590 &   1.166431 \\
ZrSiS          &   -43.368287 &   -42.910675 &   0.457612 \\
ZrSiSe         &   -41.300060 &   -41.087021 &   0.213039 \\
ZrSiTe         &   -37.549321 &   -38.490033 &   0.059288 \\
\hline
ZrGeO(ZrSiO)   &   -43.800281 &   -42.645101 &   1.155180 \\
ZrGeS          &   -41.324696 &   -40.720312 &   0.604384 \\
ZrGeSe         &   -39.401731 &   -39.076459 &   0.325272 \\
ZrGeTe         &   -37.200315 &   -37.028521 &   0.171794 \\
\hline
ZrSnO(ZrSiO)   &   -37.118496 &   -36.040903 &   1.077593 \\
ZrSnS(ZrGeS)   &   -36.269864 &   -35.743714 &   0.526150 \\
ZrSnSe(ZrGeSe) &   -35.139562 &   -34.880462 &   0.259100 \\
ZrSnTe         &   -35.395074 &   -34.708958 &   0.686116 \\
\hline
\end{tabular}
\label{table:ZrHM}
\end{table}

\begin{table}[h]
\caption{Binding energy of Hf$HM$ ($H$=Si, Ge, Sn and $M$=O, S, Se and Te) with GGA}
\centering
\begin{tabular}{lccc}
\hline\hline
              &   Bulk Energy (eV) &   Single layer Energy (eV) &   Binding Energy (eV) \\
\hline
HfSiO(HfSiS)   &   -48.066551 &   -46.724947 &   1.341604 \\
HfSiS          &   -45.837118 &   -45.424545 &   0.412576 \\
HfSiSe         &   -43.643661 &   -43.434275 &   0.209386 \\
HfSiTe         &   -41.276397 &   -41.249520 &   0.026877 \\
\hline
HfGeO(HfSiS)   &   -45.697729 &   -44.372621 &   1.325108 \\
HfGeS          &   -43.692550 &   -43.000363 &   0.692187 \\
HfGeSe         &   -41.681841 &   -41.251401 &   0.430440 \\
HfGeTe         &   -39.368634 &   -39.127598 &   0.241036 \\
\hline
HfSnO(HfSiS)   &   -39.050038 &   -37.800096 &   1.249942 \\
HfSnS(HfSiS)   &   -36.924244 &   -36.620387 &   0.303857 \\
HfSnSe(HfGeSe) &   -37.274366 &   -36.910855 &   0.363511 \\
HfSnTe(HfGeTe) &   -35.721706 &   -35.527372 &   0.194334 \\
\hline
\end{tabular}
\label{table:HfHM}
\end{table}

\begin{table}[h]
\caption{Binding energy of ZrSi$M$ ($M$=O, S, Se and Te) with LDA}
\centering
\begin{tabular}{lccc}
\hline\hline
              &   Bulk Energy (eV) &   Single layer Energy (eV) &   Binding Energy (eV) \\
\hline
ZrSiO          &   -50.057196 &   -48.554224 &   1.502862 \\
ZrSiS          &   -47.962043 &   -46.940747&    1.021296 \\
ZrSiSe        &   -45.950533 &   -45.203211 &    0.747322 \\
ZrSiTe         &   -43.032484 &   -42.695313 &   0.337171 \\
\hline
\end{tabular}
\label{table:ZrSiM}
\end{table}

\end{document}